\documentclass[aps,twocolumn,superscriptaddress,showpacs,amssymb,prl,floatfix,longbibliography]{revtex4-1}

\usepackage{graphicx,color}
\usepackage{dcolumn}
\usepackage{amsmath}
\usepackage{amssymb}
\usepackage{epstopdf}

\usepackage[]{hyperref}
\hypersetup{colorlinks=true,linkcolor=blue,citecolor=blue,urlcolor=blue,pdfpagemode=UseNone}

\newcommand{\urusi}     {URu$_2$Si$_2$}

\begin{document}

\thispagestyle{myheadings}

\title{Measurements of the NMR Knight shift tensor and nonlinear magnetization in URu$_2$Si$_2$}

\author{M. Lawson}
\author{B. T. Bush}
\author{T. Kissikov}
\author{Z. Brubaker}
\affiliation{Department of Physics, University of California, Davis, California 95616, USA}
\affiliation{Lawrence Livermore National Laboratory, Livermore, California 94550, USA}
\author{K. R. Shirer}
\affiliation{Max Planck Institute for Chemical Physics of Solids, Noethnitzer Strasse 40, D-01187 Dresden, Germany}
\author{J. R. Jeffries}
\affiliation{Lawrence Livermore National Laboratory, Livermore, California 94550, USA}
\author{S. Ran}
\author{I Jeon}
\affiliation{Department of Physics and Center for Advanced Nanoscience, University of California, San Diego, La Jolla, California 92093, USA}
\author{M. B.  Maple}
\affiliation{Department of Physics and Center for Advanced Nanoscience, University of California, San Diego, La Jolla, California 92093, USA}
\affiliation{ Materials Science and Engineering Program, University of California, San Diego, La Jolla, California 92093, USA}
\author{N. J. Curro}
\affiliation{Department of Physics, University of California, Davis, California 95616, USA}

\date{\today}
\begin{abstract}

URu$_2$Si$_2$ exhibits an anomalous peak in the nonlinear magnetic susceptibility at the hidden order transition.  In order to investigate this anomaly, we conducted direct magnetization measurements and investigated the detailed angular dependence of the $^{29}$Si nuclear magnetic resonance Knight shift tensor. We find that the nonlinear magnetization is smaller than previously reported, and the analogous nonlinear Knight shift tensor is below the detection limit.  Our results suggest that the magnitude of the anomalous peak is sample dependent.
\end{abstract}

\pacs{76.60.-k, 75.30.Mb,  75.25.Dk, 76.60.Es}



\maketitle

In recent years, there has been a growing recognition that nonlinear response functions may uncover broken symmetries that remain hidden to linear responses. In most cases, an order parameter, such as the sublattice magnetization in the case of an antiferromagnet, couples to an external field so that the linear susceptibility exhibits an anomaly at the phase transition temperature. However, the external field may not be a conjugate variable to a `hidden' order parameter, so that the linear susceptibility remains unaffected or  is only weakly modified by the onset of long-range order.  In such cases, the higher order nonlinear susceptibility terms may exhibit anomalies at the phase transition. Recent studies of the nonlinear optical response in Sr$_2$IrO$_4$  and in YBa$_2$Cu$_3$O$_7$  have revealed broken symmetries that were not evident in the linear electric susceptibilities of these materials \cite{ZhaoIridateNature2016,ZhaoYBCONature2016}. Another important example is the heavy fermion material \urusi, which exhibits a `hidden order' at low temperature that remains one of the most enduring mysteries in condensed matter physics \cite{MydoshURSreview2011}. Although several key experiments have shed new light on the nature of the hidden order phase, the order parameter remains difficult to discern \cite{Kung2015,Tonegawa2014,BoariuPRL2013,FlouquetURS2012PRL,ChatterjeeURSPRL}. In this material, the linear magnetic susceptibility exhibits only a modest change in slope at the hidden order phase transition, whereas the nonlinear component exhibits a sharp enhancement just below the transition \cite{RamirezURS1992,Miyako1991,RamirezPRL2016}. The behavior in \urusi\ is unusual, and may  reflect the Ising nature of the coupling between the hidden order and the magnetic field \cite{Chandra1994nonlinearURS}.

Nonlinear susceptibility terms generally are enhanced at second order phase transitions; however, their magnitudes are progressively smaller with increasing powers of external field.  In contrast to the linear susceptibility, the four-fold symmetry of the nonlinear term in \urusi\ is significantly enhanced at the hidden order transition, $T_0 = 17.5$ K, and has been interpreted as a consequence of a coupling between the hidden order parameter and the magnetic field  \cite{Chandra1994nonlinearURS,RamirezPRL2016}. Indeed, the jump in the nonlinear susceptibility can be related quantitatively to the jump in the specific heat at the ordering temperature, implying that the effect is an intrinsic thermodynamic phenomenon \cite{premiSCES}.
However, the microscopic origin of the non-linear susceptibility is not well understood.
Nuclear magnetic resonance (NMR) is a powerful tool to investigate local physics, and has  proven instrumental in uncovering the discrepancies between bulk and microscopic measurements of the magnetization in electronically inhomogeneous systems \cite{KoharaURSinhomogeneity,ParkDropletsNature2013}.
The Knight shift, $K$, probes the local magnetization in the crystal via a hyperfine coupling to the electron spins.  Given the large anomaly in the nonlinear magnetization observed previously \cite{RamirezURS1992}, we estimated a nonlinear Knight shift should also be present and detectable. We measured  $K$ as a function of angle, $\theta$, between the magnetic field and the crystal $c$-axis, and find that, contrary to expectation, this quantity varies  only as $\cos^2\theta$, as expected for a second-rank tensor.  In order to further investigate the discrepancy between the local and macroscopic response we measured the nonlinear magnetization in the same sample, and found that the anomalous jump at $T_0$ was reduced compared to previous reports \cite{RamirezURS1992,RamirezPRL2016}. Our results suggest that the nonlinear susceptibility observed in this sample is consistent with thermodynamic constraints, but is sample dependent.

\begin{figure}
	\centering
    \includegraphics[width=\linewidth]{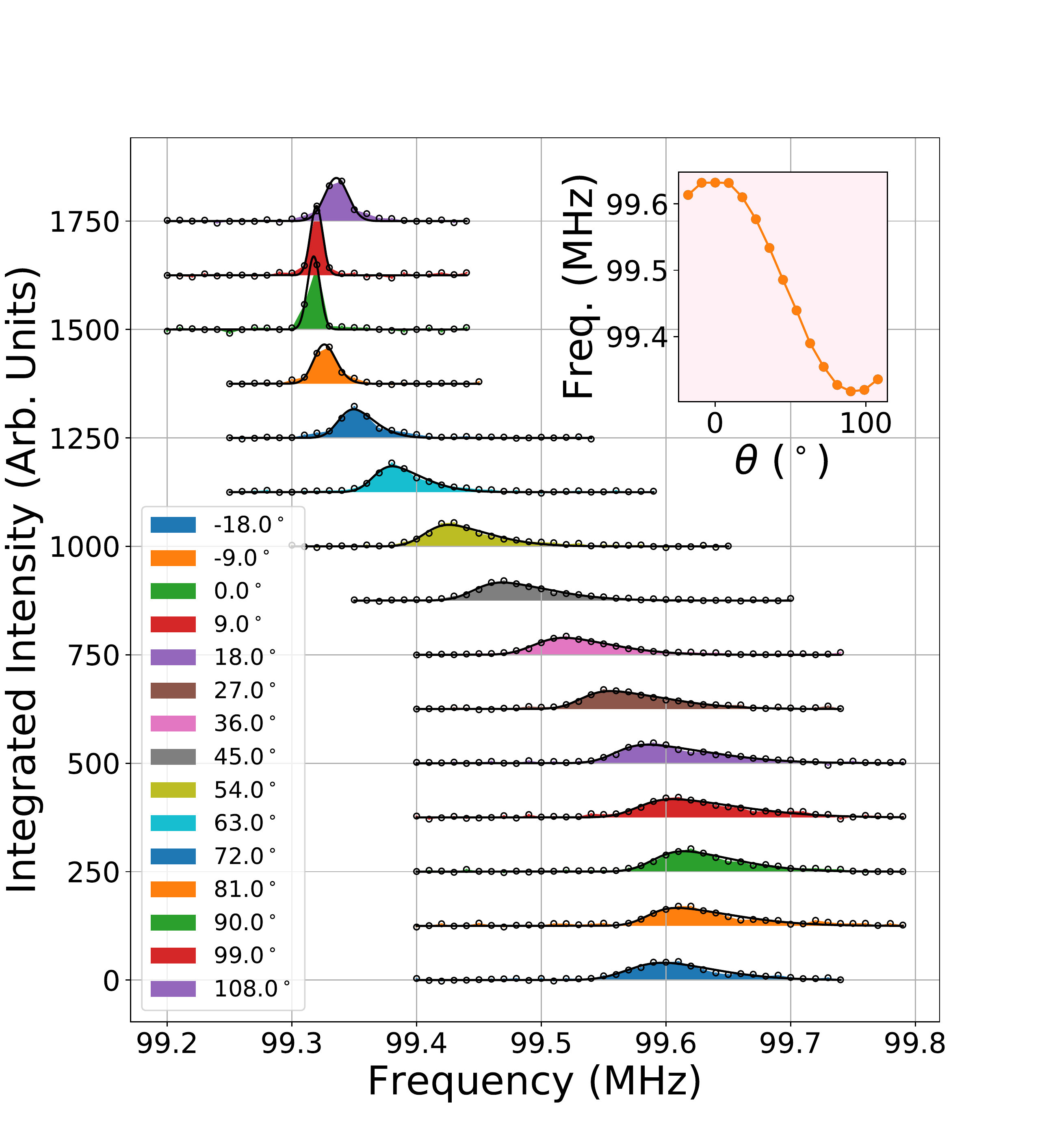}
	\caption{\label{fig:spectra} (color online)  $^{29}$Si NMR spectra of  URu$_2$Si$_{2}$ as a function of angle at 16 K. Solid black lines are fits as described in the text. The inset shows the angular dependence of the mean frequency.}
\end{figure}

\begin{figure}
	\centering
	\includegraphics[width=\linewidth]{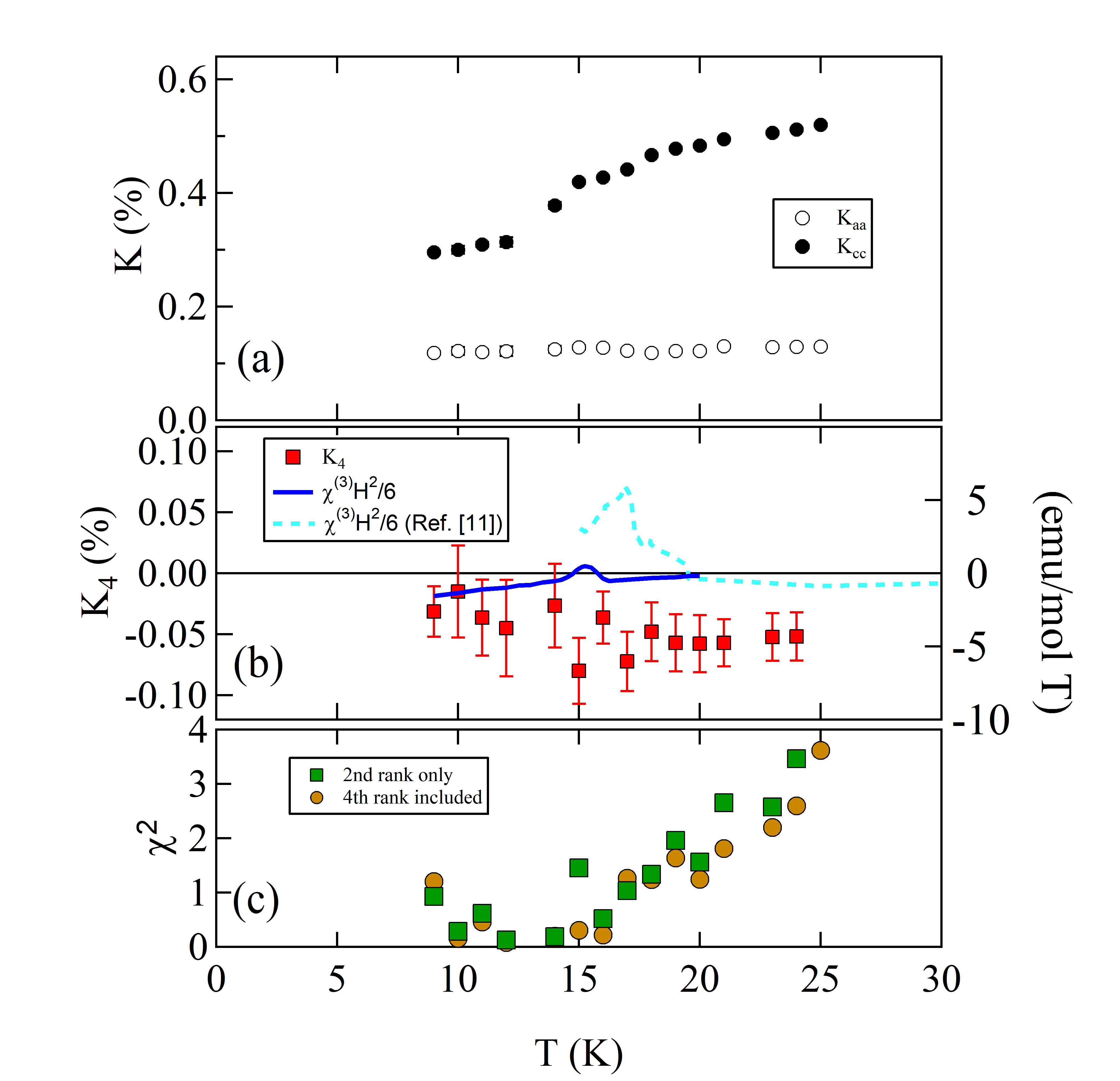}
	\caption{\label{fig:KvsTemp} (color online) (a)  Knight shift (error bars are the point sizes), (b) $K_4$ and $\chi^{(3)} H^2/6$, and  (c) the reduced $\chi^2$ both for including just the second rank, or second and fourth-rank tensors in the fits. The $\chi_3$ data in (b) (dotted line) is reproduced from \cite{RamirezPRL2016}. }
\end{figure}

A single crystal was grown by means of the Czochralski method  with isotopically-enriched $^{29}$Si ($I=1/2$), with a nominal concentration of 31\%. A thin rectangular prism of approximate dimensions 7.6 mm$\times$ 5.1 mm$\times$ 1.7 mm was cut from the growth rod, with the $c$-axis perpendicular to the main face.   The crystal was mounted in a custom-built dual-axis goniometer NMR probe \cite{Shiroka2012}, so that the crystalline $c$-axis could be rotated between $-20^{\circ}$ and $120^{\circ}$ with respect to the fixed magnetic field $\mathbf{H}_0$ of 11.7 T (see inset Fig. \ref{fig:spectra}).  The angle was measured by counting the number of turns of the drive rod, and the gear ratio is approximately 1:600, so that one rotation of the drive rod is equivalent to $0.6^{\circ}$.  At this field, the contribution of the nonlinear susceptibility is significant, so that the magnetization should exhibit a small but discernable anomaly at the hidden order transition, as shown in Fig. \ref{fig:KvsTemp} \cite{RamirezURS1992,RamirezPRL2016}.  NMR spectra were collected by summing a sequence of Carr-Purcell-Meiboom-Gill (CPMG) echoes as a function of frequency, angle, and temperature. Because both the width and the spin-echo decay time $T_2$ varied significantly with angle and temperature, the CPMG parameters were updated for each case.  Spectra were obtained by sweeping the frequency, while acquiring echo signals in both channels in quadrature detection, and utilizing computer-controlled stepper motors to rotate the goniometer orientation.  The signals were phase adjusted and the echo was integrated in a single channel. Errors in the echo integrals were determined by scaling the standard
deviation of the noise baseline by the square root of the number of integration points using the python uncertainties package \cite{pythonerrors}.
A representative data set is shown in Fig. \ref{fig:spectra} at 18 K.  These spectra were fit to a skewed Gaussian to extract the mean, $\omega_0$, and standard deviation, $\sigma$.  The fits also capture the skew asymmetry of the spectra seen in Fig. \ref{fig:spectra}. The skew arises due to a variation of the demagnetization field over the volume of the crystal, as discussed below. The shift of the resonance, $K$, is determined by the mean frequency: $K = \omega_0/\gamma H_0 -1 $, where $\gamma = 8.458$ MHz/T is the gyromagnetic ratio of the Si.

There are two important  contributions to the shift of the NMR resonance in solids: $K = K^s + K^d$, where the Knight shift $K^s$ arises due to the hyperfine coupling to the electron magnetic moments, and $K^d$ is the contribution from the demagnetization field of the sample. In general, $K^s \gg K^d$ because the hyperfine coupling between the nuclear and electron moments is large. In heavy fermions, $K^s$ can have a complicated relationship to the magnetization of the electrons due to the presence of multiple hyperfine couplings. The hyperfine Hamiltonian is: $\mathcal{H} = \mathbf{I}\cdot(A_c \mathbf{S}_c + A_f\mathbf{S}_f)$, where $\mathbf{S}_c$ is the conduction electron spin and  $\mathbf{S}_f$ is the f-electron spin, with hyperfine couplings $A_c$ and $A_f$ to the two types of spins.
In a magnetic field, the nuclei experience a hyperfine field $H_{\alpha} =  A_f M_{\alpha}/ (g\mu_B\gamma \hbar)+\Delta H_{\alpha}$ in the $\alpha$ direction, where $M_{\alpha}$ is the magnetization and $\Delta H_{\alpha}$ is proportional to $A_c-A_f$ and becomes non-zero below the coherence temperature, $T^*$, in heavy fermions \cite{ROPP2016}. However, since $A_c\approx A_f$ in \urusi, $\Delta H_{\alpha}/H_{\alpha} \lesssim 0.02$, and this term is negligible.  Relaxing this assumption, however, does not affect the conclusions, as we show below.

\begin{figure}
	\centering
	\includegraphics[width=\linewidth]{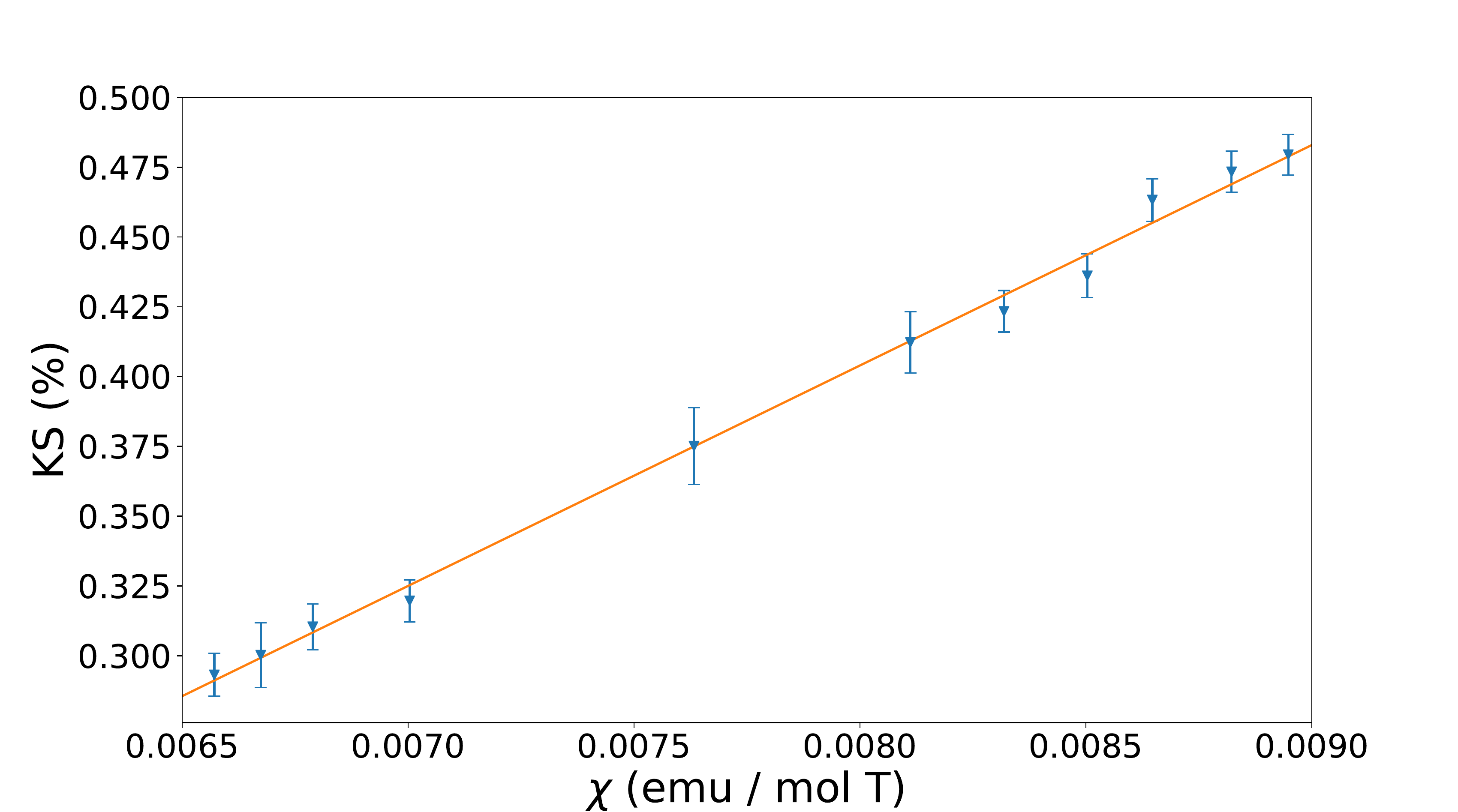}
	\caption{\label{fig:KvsChi} (color online) $K_{cc}$ versus $\chi^{(1)}_{cc}$ measured at 12 T. The solid line is linear fit with hyperfine coupling constant $A_{f} = 4.4\pm 0.1$ kOe/$\mu_B$, consistent with published data \cite{ShirerPNAS2012}. }
\end{figure}

In general, the magnetization is given by both the second-rank linear susceptibility, ${\chi^{(1)}}$,  and fourth-rank nonlinear susceptibility, ${\chi^{(3)}}$, tensors \cite{NonlinearMagnetization1984}:
\begin{equation}
{M}_\alpha = {\chi^{(1)}}_{\alpha\beta}{H}_{\beta} + \frac{1}{3!}{\chi^{(3)}}_{\alpha\beta\gamma\delta}H_{\beta}H_{\gamma}H_{\delta}+\cdots
\end{equation}
$\chi^{(1)}$ contains two independent parameters, $\chi_{aa}$ and $\chi_{cc}$. In \urusi, these are highly anisotropic and reflect the Ising-like nature of the g-factor in this material \cite{HarrisonPRL2011URS,Altarawneh2012}.  $\chi^{(3)}$ contains four independent parameters, $\chi_{aaaa}$, $\chi_{aabb}$, $\chi_{aacc}$, and $\chi_{cccc}$,  as shown by Trinh et al \cite{RamirezPRL2016}. The Knight shift thus acquires the tensor nature of the susceptibilities:  $K^s_{\alpha\beta} = (A_f/g\mu_B\gamma \hbar){\chi^{(1)}}_{\alpha\beta}$ is the usual second-rank Knight shift tensor, and $K^s_{\alpha\beta\gamma\delta} = (A_f/g\mu_B\gamma \hbar){\chi^{(3)}}_{\alpha\beta\gamma\delta}$ is the fourth-rank Knight shift tensor.  The Knight shift is then:
\begin{equation}
\label{eqn:Kvstheta}
K^s(\theta) = K_0 + K_2\cos^2(\theta) + K_4\cos^4(\theta),
\end{equation}
where $\theta$ is the angle between the external field $\mathbf{H}_0$ and the $c$-axis. The coefficients are given in Appendix A.  Importantly, $K_4$ depends only on $\chi^{(3)}$, so it vanishes in the absence of nonlinear contributions.  Note we assume that the hyperfine coupling tensor is isotropic \cite{kohoriURu2Si2}, however, any anisotropies in the hyperfine coupling tensor would only modify the values of $K_0$ and $K_2$, but not affect $K_4$.

\begin{figure}[h!]
	\centering
    \includegraphics[width=\linewidth]{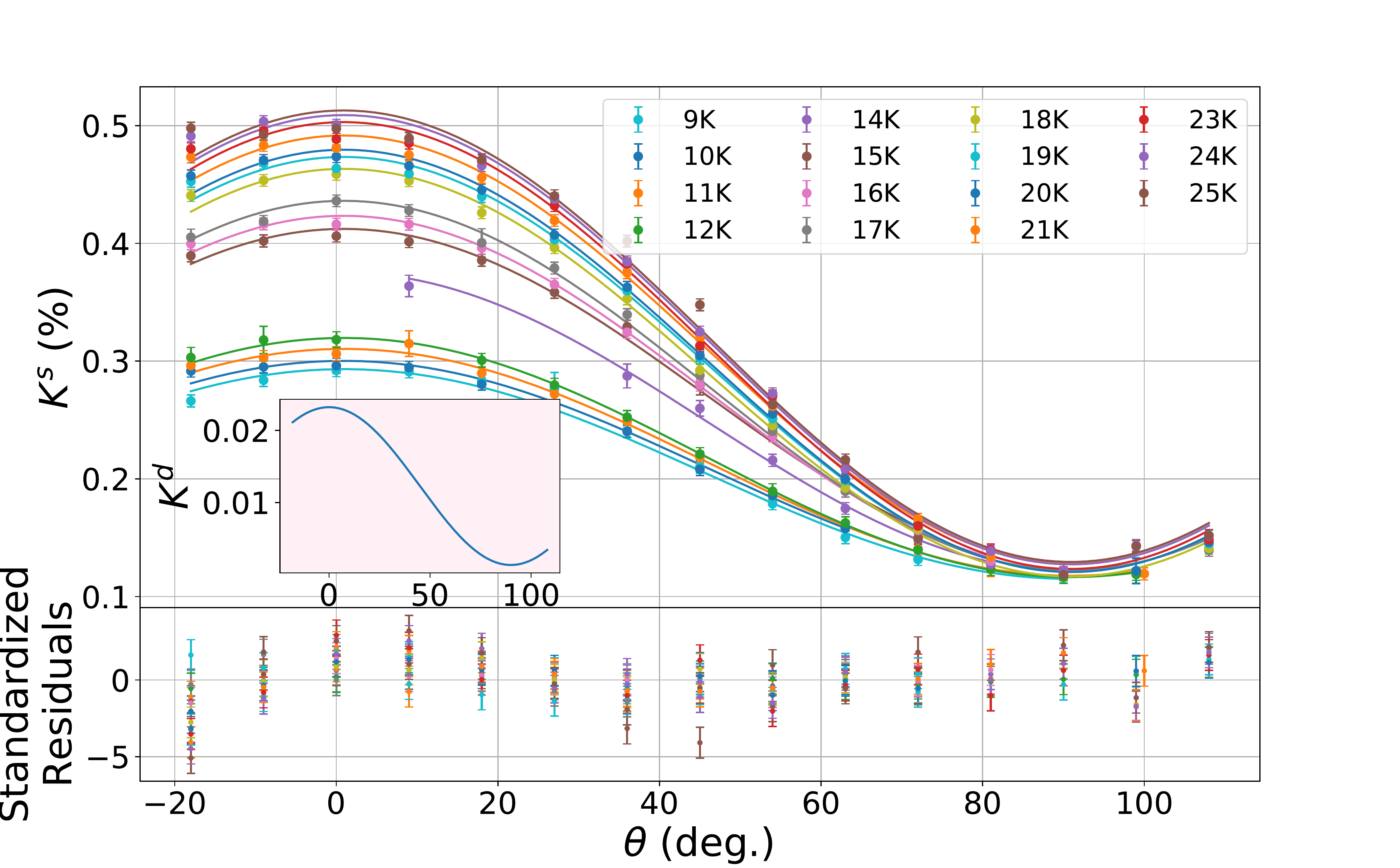}
	\includegraphics[width=\linewidth]{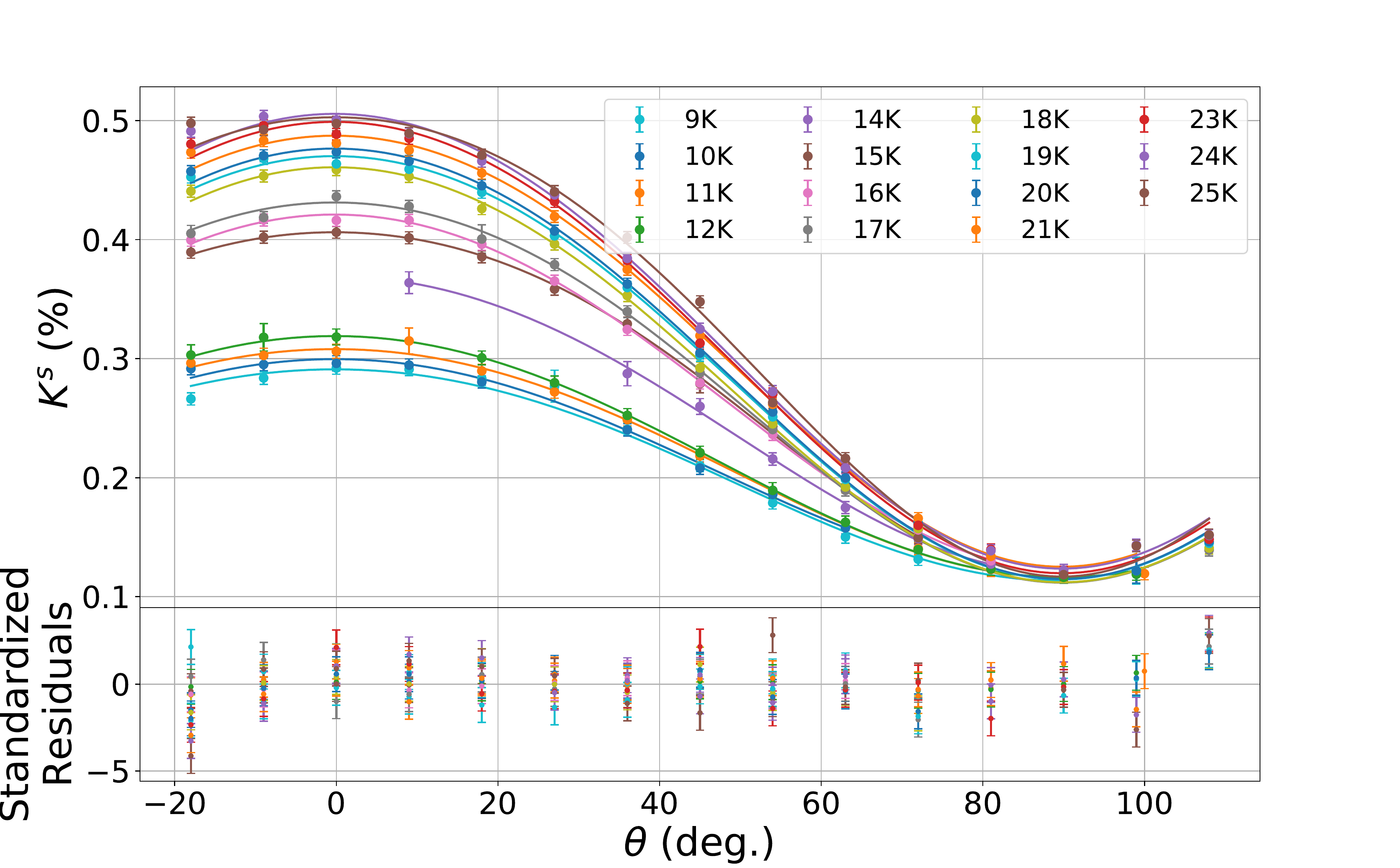}
    \caption{\label{fig:Ksummary} (color online)  Knight shift, $K^s$, versus angle and temperature with (upper) and without (lower) including the fourth-rank tensor contribution.  The solid lines are fits as described in the text, and residuals are shown below for each set of fits. The error bars were computed by summing in quadrature the fitting parameter uncertainty for the spectral fits given the measured echo integral errors, a 10\% uncertainty in both the demagnetization factor and in the magnetic susceptibility at each angle, as discussed in the text. The inset shows the demagnetization shift as a function of angle, as discussed in the text.}
\end{figure}

\begin{figure}[!tb]
	\centering
    \includegraphics[width=\linewidth]{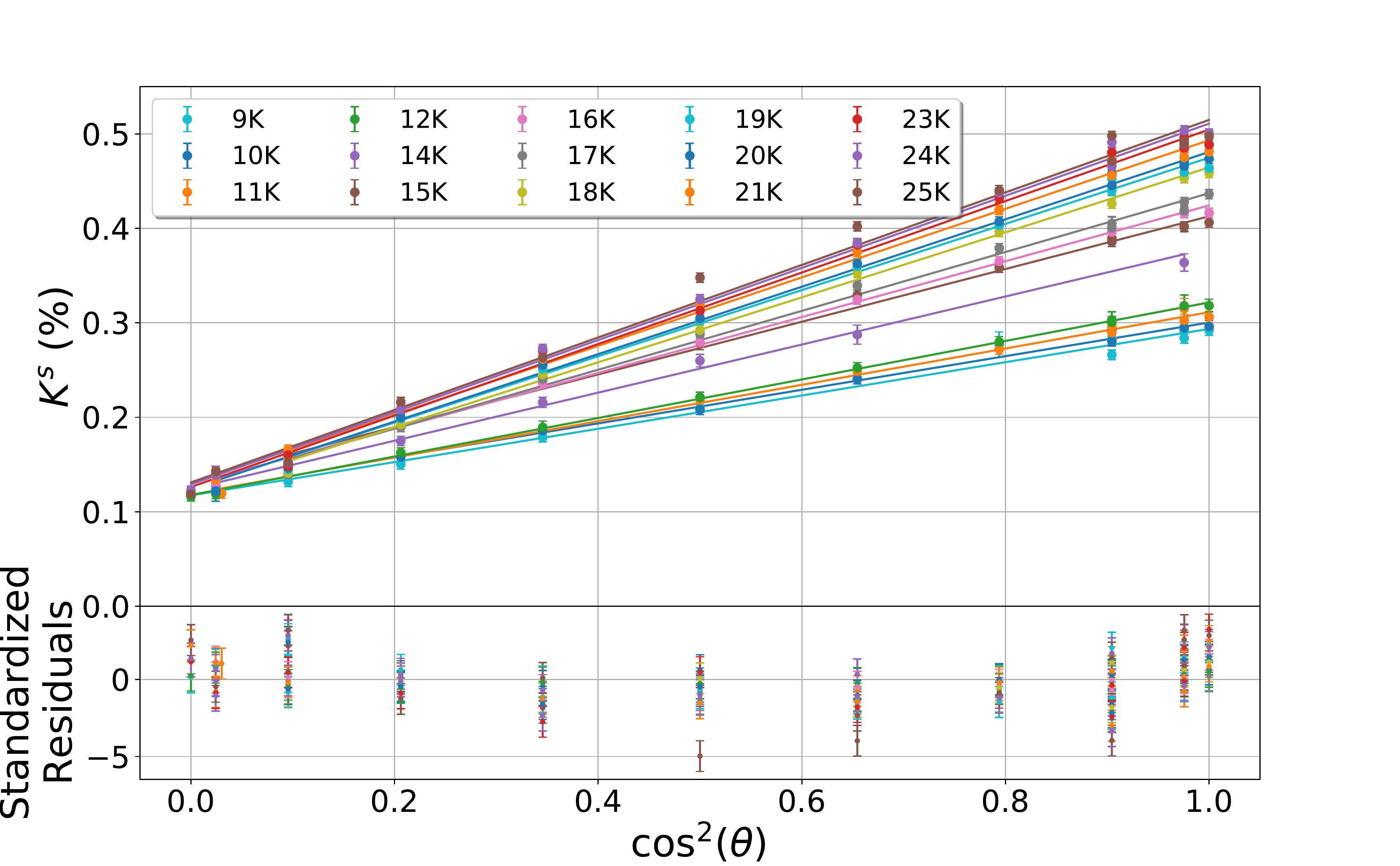}
	\caption{\label{fig:KvsCos2} (color online)  Knight shift versus $\cos^2\theta$ and temperature. The solid lines are linear fits to the data, and the residuals are shown below.  The error bars are the same as those shown in Fig. \ref{fig:Ksummary}.}
\end{figure}

It is important to also account for the angular dependence of the demagnetization shift, $K^d=|\mathbf{B}|/H_0 -1$.   Fig. \ref{fig:Ksummary} show $K^s = K - K^d$, in which the angular dependence of the demagnetization field has been subtracted, as described in Appendix B and shown in the inset.  The solid lines in Fig. \ref{fig:Ksummary} are fits to Eq. \ref{eqn:Kvstheta}, in which $K_4$ is either held at zero or allowed to float.  Fig. \ref{fig:KvsTemp} shows how these fit parameters vary as as a function of temperature, as well as the reduced $\chi^2$ value for the angular-dependent fits.  For the fits without the $K_4$ term, the residuals exhibit tiny  deviations from zero that are slightly improved by including the $K_4$ term.  However, the reduced $\chi^2$ values exhibit no statistically-significant improvement in the goodness-of-fit.
The absence of a $K_4$ term is also highlighted in the linear dependence seen in Fig. \ref{fig:KvsCos2}, which shows $K^s$ versus $\cos^2\theta$.

\begin{figure}
	\centering
	\includegraphics[width=\linewidth]{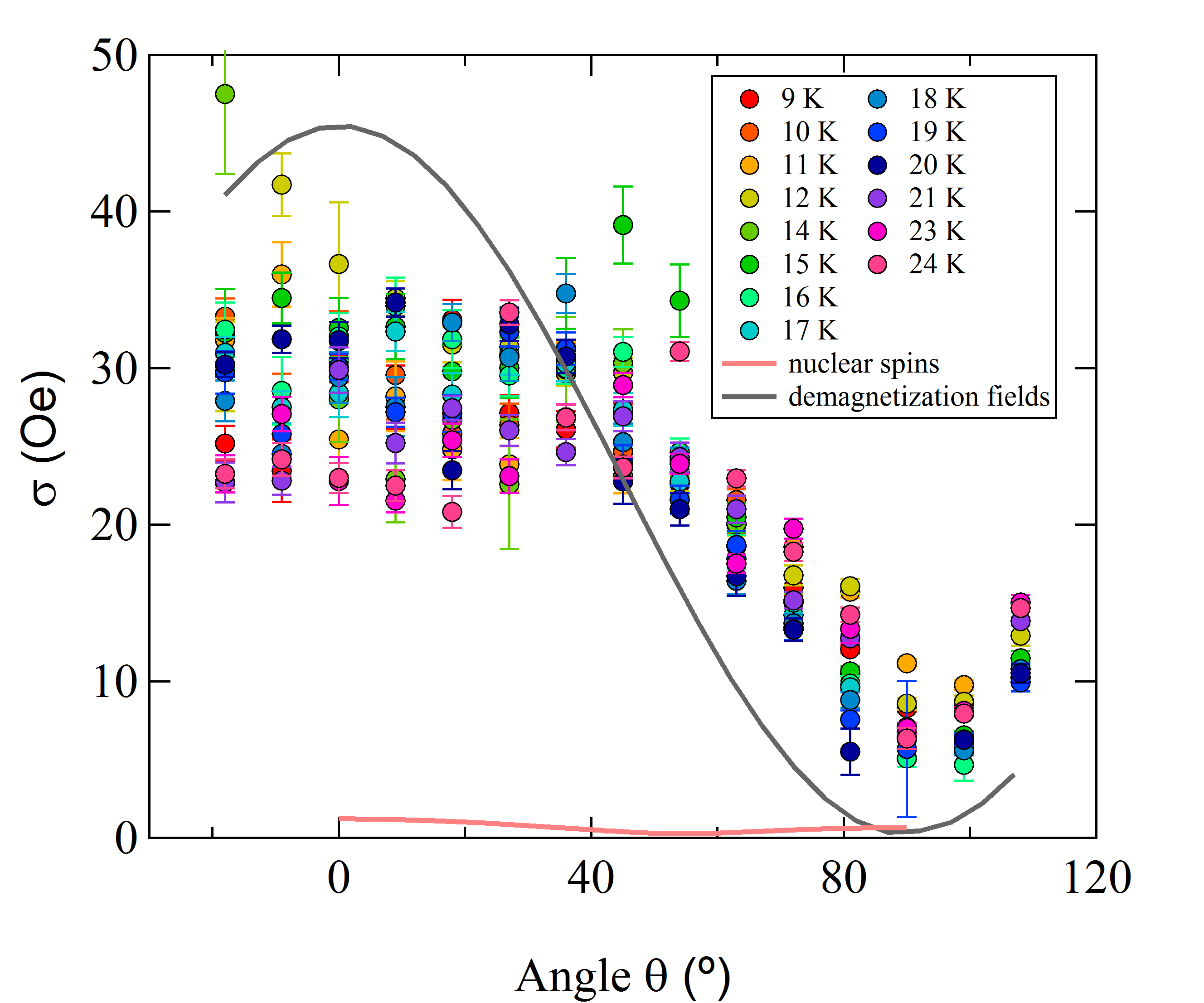}
    	\caption{\label{fig:std} (color online)  Standard deviation of the lineshape versus angle and temperature. The red solid line is the second moment of the nuclear dipole spin fields, and the solid gray line is the second moment of the demagnetization fields, as described in the text.}
\end{figure}

The Knight shift was determined solely based on the mean values extracted to from the fits to spectra shown in Fig. \ref{fig:spectra}, however the spectra also exhibit a pronounced variation in linewidth as a function of angle.  The standard deviation of the spectra, $\sigma$, is shown in Fig. \ref{fig:std}.  $\sigma$ is maximum for $\theta =0$ ($\mathbf{H}_0~||~[001]$) and minimum for $\theta = 90^{\circ}$ ($\mathbf{H}_0~||~ [100]$).  The resonance can be broadened by  (i) the variation of local fields due to the neighboring dipolar fields of other $^{29}$Si nuclei in the lattice, (ii) the dipolar field of the $^{101}$Ru neighbors, (iii) electronic inhomogeneity within the sample, and (iv) variation of the local demagnetization field within the sample itself.    The calculated broadening due to the nuclear dipole-dipole coupling is described in Appendix C and is shown in Fig. \ref{fig:std} as a function of $\theta$ (red line). This contribution is only 2-3 Oe, an order of magnitude smaller than the experimental values. It is clear, therefore, that the dominant contribution to the linewidth arises either from (iii) or (iv).  We estimate the contribution from variations of the demagnetization field over the volume of the sample, as a function of field orientation. The standard deviation of the local field distribution due to the inhomogeneous demagnetization field is shown in Fig. \ref{fig:std} as the solid black line.  Although the agreement with the measured values is not exact, the order of magnitude and the trend with angle agrees well.
We note further that the skew observed in the spectra is also clear in the histogram of the computed local fields, with a tail extending to higher fields for angles close to $0^{\circ}$, in agreement with the data shown in Fig. \ref{fig:spectra}.

The temperature dependence of $K_{cc}$, $K_{aa}$ and $K_4$ are shown in Figs. \ref{fig:KvsTemp}(a) and (b), where $K_{aa} = K_0$ and $K_{cc} = K_0 + K_2$.  $K_{cc}$ and $K_{aa}$ behave similarly to the bulk linear susceptibilities, $\chi_{cc}$ and $\chi_{aa}$, and $K_{cc}$ exhibits a slight change in slope at $T_0$. $K_{cc}$ is linearly proportional to $\chi_{cc}$, as seen in Fig. \ref{fig:KvsChi}. The anisotropy of these Knight shifts is consistent with previous measurements of the Ising anisotropy observed in this material \cite{RamirezPRL2016}.  The nonlinear component, $K_4$, is essentially temperature-independent and close to zero. This non-zero average value of $-0.007\pm0.003\%$ may be the result of a systematic error, possibly due to a misalignment of less than 1$^{\circ}$ towards the axis of rotation.  The angular fits shown in Fig. \ref{fig:Ksummary} yield an offset of $\sim 1.4^{\circ}$, consistent with this scenario.  The nonlinear susceptibility reported in Ref. \cite{RamirezPRL2016} is also shown in Fig. \ref{fig:KvsTemp}(b), as well as that measured on  a crystal from the same batch as the Knight shift.  The measured jump $\Delta\chi_3$ we observe is only 16\% of that reported in Ref. \cite{RamirezPRL2016} (and only 1.7\% of that reported in Ref.  \cite{RamirezURS1992}).   The jump in the specific heat, $\Delta C_V/T = 146$ mJ/mol$\cdot$K$^2$, and the change in slope of the linear susceptibility at $T_{0}$, $\Delta ({d\chi_1}/{dT}) = 2.558$ emu/(mol$\cdot$T$\cdot$K) of the current sample yield the thermodynamic relation $\frac{\Delta C_V}{T}\left(\Delta\chi_3\right)\left(\Delta{d\chi_1}/{dT}\right)^{-2}\approx 1.6$, which the same order of magnitude as the theoretical value of 3 \cite{premiSCES}.

In summary, we have measured the Knight shift through the hidden order transition in \urusi\ and found no evidence for a fourth-rank tensor that would give rise to a four-fold symmetry as a function of angle.  Our results differ from recent measurements of the bulk magnetization, in which the nonlinear susceptibility exhibits a sharp anomaly at the hidden order transition and displays a $\cos^4\theta$ angular variation. Direct measurements of the nonlinear susceptibility in the current sample, however, reveal a nonlinear anomaly that is approximately ten times smaller than previously observed, which would not be detectable within the precision of our Knight shift measurements.  The anomalous jump in the nonlinear susceptibility is related to the field-dependence of $T_0(H)$, and different samples may exhibit the same transition temperatures but with different anomalous magnetic behavior.   The isotopically-enriched sample used in this study exhibits a $T_0=16.4$ K, but the size of the thermodynamic anomalies at $T_0$ are consistent with thermodynamic predictions.  However, unlike the nonlinear optical studies in Sr$_2$RuO$_4$ and YBa$_2$Cu$_3$O$_7$, the anomalous nonlinear magnetization in URu$_2$Si$_2$ does not reveal fundamental new information about the broken symmetry in this material, but rather is a natural consequence of the field-dependence of $T_0$.

Nonlinear magnetization has also been observed by torque magnetometry in \urusi\ as a four-fold $\cos(4\phi)$ variation for field oriented in the basal plane \cite{Okazaki2011}.  Knight shift measurements do not reveal any in-plane angular variations, but do find a four-fold variation of the NMR linewidth (in addition to a two-fold variation) \cite{KambeURS2013}.  This variation may reflect a finite $\chi^{(3)}$ term \cite{NonlinearMagnetization1984,RamirezPRL2016}, however the magnitude of the measured effect is independent of temperature and has been interpreted as originating from local disorder \cite{KambeURS2015PRB}. Our finding that the magnitude of the temperature dependence of $\chi^{(3)}$ is sample dependent suggests a possible role of local disorder.  Further studies are necessary to determine to what extent impurities and disorder can affect the nonlinear magnetization in \urusi.

\begin{acknowledgements}
We thank G. Kotliar, A. Ramirez, P. Chandra, P. Coleman, and G. Blumberg for enlightening discussions, and K. DeLong, D.Hemer and P. Klavins for assistance in the laboratory. Work at UC Davis was supported by the NSF under Grant No.\ DMR-1506961 and the National Nuclear Security Administration under the Stewardship Science Academic Alliances program through DOE Research Grant \#DOE DE-FG52-09NA29464. Research at UCSD was supported by the U. S. Department of Energy, Office of Basic Energy Sciences, Division of Materials Sciences and Engineering under Grant No. DE-FG02- 04ER46105. A portion of this work was completed at the Aspen Center for Physics, which is supported by the National Science Foundation grant PHY-1066293. Lawrence Livermore National Laboratory is operated by Lawrence Livermore National Security, LLC, for the U.S. Department of Energy, National Nuclear Security Administration under Contract DE-AC52-07NA27344.

\end{acknowledgements}

\appendix
\section{Appendix A: Nonlinear susceptibility tensor}
Trinh et al. \cite{RamirezPRL2016} show that the anisotropic susceptibility is well described by $\chi_3 = \chi _{\text{aaaa}} + (\chi_{\text{cccc}}+\chi_{\text{aaaa}} - 6 \chi_{\text{aacc}})\cos^4\theta$.  The Knight shift coefficients in  Eq. \ref{eqn:Kvstheta}  are thus given by:
\begin{eqnarray}
K_0 &=& \frac{A_f}{g\mu_B\gamma \hbar} \left(\chi _{\text{aa}}+\frac{H_0^2}{6}\chi _{\text{aaaa}}\right)\\
K_2 &=& \frac{A_f}{g\mu_B\gamma \hbar}  \left(\chi _{\text{cc}}- \chi _{\text{aa}} \right) \\
K_4 &=&\frac{A_f}{6g\mu_B\gamma \hbar}  H_0^2 \left(\chi_{\text{cccc}}+\chi_{\text{aaaa}} - 6 \chi_{\text{aacc}}\right).
\end{eqnarray}

\section{Appendix B: Demagnetization Field}

The contribution from the demagnetization field can be computed numerically using the measured susceptibility tensor and the dimensions of the crystal. We computed the demagnetization tensor, $\mathbb{D}(\mathbf{r})$, for points within the sample using the magnetic scalar potential for a 3D uniform set  of 71,680 points within a rectangular prism of the same dimensions as the sample.
The magnetic field within the sample is then given by: $\mathbf{B}(\mathbf{r})=\left[\Gamma(\mathbf{r})+4\pi \chi\cdot\Gamma(\mathbf{r})\right]\mathbf{H}_0$, where $\Gamma(\mathbf{r}) =\left(\mathbb{I} - \mathbb{D}(\mathbf{r})\cdot\chi\right)^{-1}$. We consider here only the linear susceptibility, $\chi = \chi^{(1)}$, which nevertheless is anisotropic. We computed $\mathbf{B}(\mathbf{r})$ for various field orientations and determined $K_d(\theta) = 0.00012 + 0.00011\cos(2\theta)$, which is an order of magnitude less than the measured $K$.

\section{Appendix C: Nuclear Dipole-Dipole Coupling}

The dipole-dipole second moment is given by:
\begin{equation}
\sigma_{dip}^2 = \frac{3}{4}\gamma^4\hbar^2 I(I+1)P\sum_k \frac{1 - 3\cos^2\theta_{k}}{r_{k}^6}
\end{equation}
where $\mathbf{r}_k$ is the vector between the Si sites in the lattice,  $\theta_k$ is the angle between $\mathbf{H}_0$ and $\mathbf{r}_k$, and $P \approx 0.40$ is the abundance of the $^{29}$Si isotope.

\bibliography{CurroBibliography}

\end{document}